\documentclass[hidelinks,12pt]{article}

\usepackage{arxiv}

\usepackage[utf8]{inputenc} 
\usepackage[T1]{fontenc}    
\usepackage{hyperref}       
\usepackage{url}            
\usepackage{booktabs}       
\usepackage{amsmath,amsfonts,amssymb,amsthm,wasysym}       
\usepackage{nicefrac}       
\usepackage{microtype}      
\usepackage{cleveref}       
\usepackage{graphicx}
\usepackage{natbib}
\usepackage{doi}
\usepackage{physics}

\usepackage{caption}
\numberwithin{equation}{section}
\setcitestyle{square}

\title{Poiseuille flow for a simplified pseudoplastic rheology}

\author{Chris Reese \\
	Department of Physics\\
	Lewis and Clark Community College\\
	Godfrey, IL   62035\\
	\texttt{ccreese@lc.edu}}

\renewcommand{\headeright}{Pseudoplastic Poiseuille flow}
\renewcommand{\undertitle}{Pseudoplastic Poiseuille flow}
\renewcommand{\shorttitle}{}

\hypersetup{
pdftitle={seudoplastic Poiseuille flow},
pdfsubject={},
pdfauthor={Chris Reese},
pdfkeywords={}}

\begin{document}

\maketitle

\begin{abstract}

Poiseuille flow in cylindrical and planar geometries with a simplified, pseudoplastic (shear thinning) rheology 
characterized by constant viscosity plateaus above and below a transition strain rate is considered.  
Analytical, steady state solutions for velocity profile and volume flux are formulated.  Transient flow development is
addressed numerically and compared to the theory in the steady state limit.  Stationary flow is approached
after the momentum diffusion timescale based on the spatially dominant kinematic viscosity.  For 
large viscosity ratio and shear thinning region confined near the domain boundary, 
velocity distributions are quasi-plug like with large boundary to interior strain rate ratio.

\end{abstract}

\section{Introduction}

Due to analytical complexity, instances of closed form solutions for Poiseuille flow 
with non-Newtonian rheologies are limited. 
While the Ostwald–de Waele power-law fluid admits solutions in planar and cylindrical geometries, 
the rheological model suffers from lack of effective viscosity limits at extreme strain rates.
A truncated power law model alleviates this drawback \citep{lavrov15} and can be adopted
to fit Carreau fluid characteristics \citep{wrobel20}.
\cite{nouar09}
demonstrate the existence of stable solutions for combined plane Couette-Poiseuille flow 
of a shear thinning Carreau fluid for certain rheological parameters.  
Analytical Poiseuille flow solutions have also been obtained for a number of rheological models in the
context of the effects of boundary slip condition \citep{ferras12}. 
Volume flow rate in layer and pipe geometries can be calculated for several generalized Newtonian rheologies
utilizing both a variational principle and the so-called Weissenberg, Rabinowitsch, Mooney and Schofield method 
\citep{sochi14,sochi15a,sochi15b,sochi16}.  
\cite{griffiths20} addresses channel flow for both shear-thinning and shear-thickening Carreau fluids finding
exact solutions for particular rheological parameters.
          
In this work, a simplified pseudoplastic rheology \citep{james21} is adopted to investigate planar
and cylindrical Poiseuille flow.   The generalized Newtonian viscosity (\S2) is a shear thinning, 
three parameter function characterized by small and large strain rate viscosity 
plateaus and a rheological transition stress.  Analytical, steady state solutions for velocity 
profile and volume rate are developed in \S3.   \S4 presents numerical results for 
flow development from a quiescent initial condition and approach to steady state.   A discussion 
of results is presented in \S5.  The appendix (\S6) benchmarks the numerical method 
utilized in the study against analytical solutions for isoviscous, transient flow.

\section {Rheological model}  
\label{sec:rheo}

For a generalized Newtonian rheology, effective viscosity is a function of strain rate.  The deviatoric stress tensor
\begin{equation}
\boldsymbol{\tau} = 2 \, \eta_{\rm eff}(\dot{\varepsilon}) \, \boldsymbol{\dot{\varepsilon}} \; ,
\end{equation}
where
\begin{equation}
\boldsymbol{\dot{\varepsilon}} = \frac{1}{2} \left[ \grad \boldsymbol{u} + \grad \boldsymbol{u}^{\rm T} \right] \; ,
\end{equation} 
is the strain rate tensor and  $\boldsymbol{u}$ is fluid velocity.  The strain rate and stress invariants 
\begin{equation}
\dot{\varepsilon}  =  \left[ \frac{1}{2}  \; \Tr \left( \boldsymbol{\dot{\varepsilon}} \boldsymbol{\dot{\varepsilon}}^{\rm T} \right) \right]^{1/2} \,, \qquad
\tau   =  \left[ \frac{1}{2}  \; \Tr \left( \boldsymbol{\tau} \boldsymbol{\tau}^{\rm T} \right) \right]^{1/2} \; .
\end{equation} 

\begin{figure}[ht!]
\centering
\includegraphics{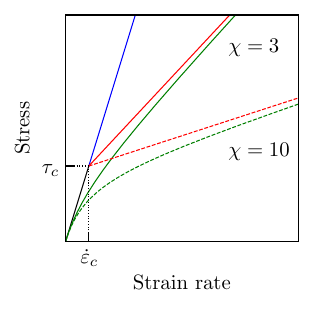}
\hspace{0.25in}
\includegraphics{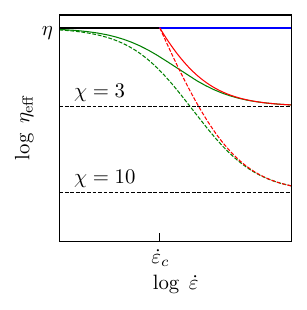}
\caption{A qualitative comparison of various rheologies.  (left) The stress invariant versus strain invariant for isoviscous (blue), simplified pseudoplastic (red), and Cross fluids (green).  All fluids have the same small strain rate viscosity.  The simplified pseduoplastic and Cross fluids approach the same asymptotic large strain rate viscosity. The parameter $\chi=\eta/\mu$ is the ratio of small strain rate to large strain rate viscosities.  Solid and dashed curves correspond to $\chi$ = 3 and 10, respectively.  (right) Effective viscosities of the rheological models.}
\label{rheo} 
\end{figure}

The piecewise linear approximation to pseudoplastic rheology \citep{james21} adopted in this study is
\begin{equation}
\eta_{\rm eff}(\dot{\varepsilon}) = \begin{cases}
\eta  & \dot{\varepsilon}  \le   \dot{\varepsilon}_{\rm c} \\
 \dfrac{\tau_c}{2  \dot{\varepsilon} } \left(1 - \dfrac{\mu}{\eta} \right) + \mu &  \dot{\varepsilon}  \ge   \dot{\varepsilon}_{\rm c} \\
\end{cases}
\label{eq:rheo}
\end{equation}
where $\dot{\varepsilon}_{\rm c}$ is the transition strain rate for onset of shear thinning,  $\tau_c = 2\, \eta \, \dot{\varepsilon}_{\rm c}$  is the transition stress invariant, $\eta$ is the low strain rate viscosity, and $\mu$ is the high strain rate viscosity.  This rheological model is compared to the isoviscous case and a Cross fluid in Fig.\ \ref{rheo}.

\section{Steady state}

For incompressible fluid flow driven solely by a pressure gradient, the Cauchy momentum equation is
\begin{equation}
 - \grad p + \grad \cdot \boldsymbol{\tau}  = \rho \; \left[\frac{\partial \boldsymbol{u}}{\partial t} + \boldsymbol{u} \cdot \grad \boldsymbol{u} \right] 
 \label{cauchy}
\end{equation}
where $p$ is pressure and $\rho$ is fluid density.  In this section, planar and cylindrical steady state 
solutions are developed for the generalized Newtonian viscosity described in \S2. 

\subsection{Planar geometry}

Consider a plane layer perpendicular to the $y$-axis and bounded by no slip surfaces at $y=0$ and $y = a$. Flow is driven by a uniform pressure gradient, $p(x) = - \nabla p \,  x$ with $\nabla p > 0$.  In this case, the steady state velocity $\boldsymbol{u} = u(y) \, \hat{\boldsymbol{i}}$ and is symmetric about the layer mid-plane.  That is, $u(y)$ is maximum at the stationary point $y=a/2$ and decreases to zero at the boundaries.   

\begin{figure}[ht!]
\centering
\includegraphics{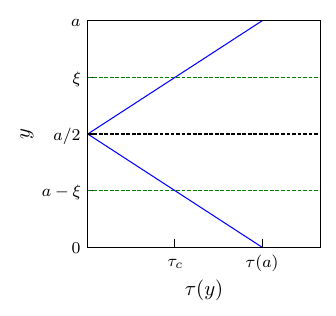}
\hspace{-0.25in}
\includegraphics{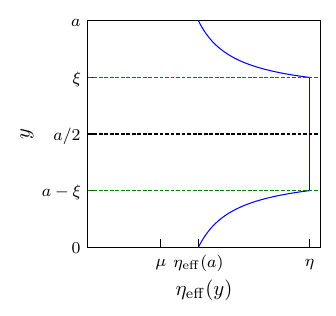}
\hspace{-0.25in}
\includegraphics{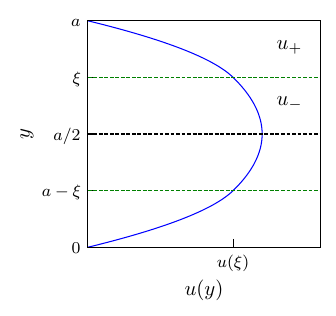}
\caption{An example planar solution with $\chi=3$ and $\xi=0.75$.  (left) Stress invariant (blue) is symmetric about the mid-layer (black dashed) increasing linearly from zero to maxima  $\nabla p \, a/2$ at the boundaries.  For $\tau_c < \tau(a)$, shear thinning regions exist for $y>\xi$ and $y < a-\xi$ (green dashed).  (middle) Effective viscosity (blue). (right) Velocity profile (blue).  In the upper half layer the velocities above and below the shear thinning transition are denoted by $u_+$ and $u_-$, respectively.  }
\label{diag} 
\end{figure}

The non-zero components of the stress tensor
\begin{equation}
\tau_{xy} = 2 \, \eta_{\rm eff}(\dot{\varepsilon}) \, \dot{\varepsilon}_{xy} = 2 \, \eta_{\rm eff} (\dot{\varepsilon}) \, \left(\frac{1}{2} \frac{d u}{d y}\right) = \eta_{\rm eff} (\dot{\varepsilon}) \, \frac{d u}{d y}  
\end{equation}
The momentum equation reduces to
\begin{equation}
\frac{dp}{dx} = - \nabla p = \frac{d \tau_{xy}}{dy \hfill} 
\end{equation}
Integrating and applying the symmetry condition $\tau_{xy} = 0$ at $y=a/2$ yields the shear stress
\begin{equation}
\tau_{xy} = - \nabla p \left( y - \frac{a}{2} \right)
\label{tauxy}
\end{equation}
The stress invariant $\tau(y) = \nabla p \left| y-a/2 \right|$ increases linearly with distance from the mid-plane with maximum stress invariant at the no-slip layer boundaries, 
$\tau(0) = \tau(a) = \nabla p \; a/2$ (Fig.\ \ref{diag}).  For transition stress $\tau_c \ge \tau(a)$, the entire layer is in the high viscosity rheological regime. 
For $\tau_c < \tau(a)$, there are two $y$ positions where $\tau = \tau_c$.  Denoting the upper half-plane position where this occurs as $\xi$ gives 
\begin{equation}
\xi = \frac{a}{2} + \frac{\tau_c}{\nabla p} \; .
\label{xiplan}
\end{equation}
In the lower half-plane, the position below which the stress invariant exceeds $\tau_c$, is $y=a-\xi$ (Fig.\ \ref{diag}).   

\begin{figure}[ht!]
\centering
\includegraphics{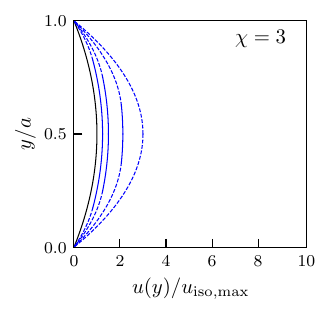}
\hspace{-0.25in}
\includegraphics{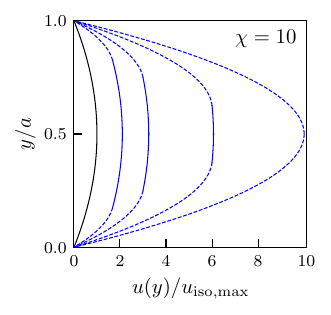}
\hspace{-0.25in}
\includegraphics{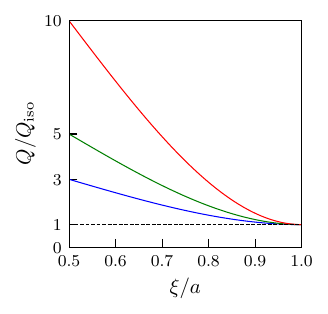}
\caption{Velocity profile dependence on viscosity ratio $\chi$ and rheological transition level $\xi$.  In this graph, the coordinate $y$ is non-dimensionalized by layer thickness $a$ and the velocity has been scaled by the maximum isoviscous velocity $u_{\rm iso,max} = (\nabla p \, a^2)/(8 \eta)$.  (left) The isoviscous case (solid black line) is compared to pseudoplastic solutions (blue lines) for $\chi=3$ and $\xi$ = (0.825, 0.750, 0.625, 0.500).  Layer regions where $\tau > \tau_c$ are denoted by dashed blue lines.  (center) Velocity profiles for $\chi$ = 10 and the same values of $\xi$.  (right) Volume rate (per unit length in the $z$ direction) as a function of rheological transition position $\xi$. The isoviscous case (dashed black line) is compared to pseudoplastic solutions for $\chi$ = (3,5,10) shown in (blue, green, red) respectively.}
\label{fig2} 
\end{figure}

The velocity profile $u(y)$ is formulated for the upper half-plane, $a/2 \le y \le a$.  Reflection symmetry about the mid-plane gives the velocity in the lower half-plane.   
Let $u_-$ be the velocity for $a/2 \le y < \xi$ where $\tau<\tau_c$ and $\eta_{\rm eff} = \eta$.  Then, Eq.\ (\ref{tauxy}) becomes
\begin{equation}
\eta \frac{d u_-}{d y \hfill} = - \nabla p \left(y-\frac{a}{2} \right)
\end{equation}  
Integrating
\begin{equation}
u_- = \frac{\nabla p \, a^2}{2 \eta} \, \frac{y}{a} \left(1 - \frac{y}{a} \right) + k_1
\end{equation}
where $k_1$ is a constant.
Let $u_+$ be the velocity above the viscosity transition level, i.e., for $\xi \le y \le a$.  Upon substituting for the strain rate invariant, $\dot{\varepsilon} = 1/2 \, |du_+/dy|$ in the effective viscosity, Eq.\ (\ref{tauxy}) yields,
\begin{equation}
\frac{\eta}{\chi} \left[ \frac{\tau_c}{\eta \left|\dfrac{d u_+}{d y \hfill} \right|} \left(\chi-1\right)+1\right] \dfrac{d u_+}{d y \hfill} = - \nabla p \left(y-\frac{a}{2} \right)
\end{equation} 
where $\chi = \eta/\mu$ is the high to low viscosity ratio.  
In the upper half plane, the velocity gradient is monotonic and less than zero, i.e., $ |du_+/dy| = - du_+/dy$. Substituting, eliminating $\tau_c$ using Eq.\ (\ref{xiplan}), integrating, and applying the no slip boundary condition $u_+(a) = 0$ yields
\begin{equation}
u_+ = \frac{\nabla p \, a^2}{2 \eta} \left[\chi \frac{y}{a} \left(1 - \frac{y}{a} \right) - 2 (\chi-1) \left(\frac{\xi}{a}-\frac{1}{2}\right) \left(1-\frac{y}{a}\right)  \right] 
\end{equation}
Matching solutions, $u_-(\xi) = u_+(\xi)$ yields the constant $k_1$ and
\begin{equation}
u_- = \frac{\nabla p \, a^2}{2 \eta} \left[\frac{y}{a} \left(1 - \frac{y}{a} \right) + (\chi-1) \left(1-\frac{\xi}{a}\right)^2 \right] 
\end{equation}
Velocity profiles for some values of $\chi$ and $\xi$ are shown in Fig.\ (\ref{fig2}).

Various limits can be checked against isoviscous cases.  For  $\eta = \mu$, $\chi = 1$ and
$u_+=u_-=u_{\rm iso}$ where the isoviscous solution
\begin{equation}
u_{\rm iso} = \frac{\nabla p \, a^2}{2 \eta} \, \frac{y}{a} \left(1 - \frac{y}{a} \right)
\end{equation}
When $\tau_c = \tau(a)$, $\xi=a$, the entire layer is in the high viscosity regime, and $u_-=u_{\rm iso}$.  In the large strain rate limit, $\tau_c \ll \tau(a)$ and $\xi \rightarrow a/2^+$.  Apart from a narrow, small strain rate sublayer around $y=a/2$, the rest of the layer is in the low viscosity regime.    In this case, $u_+ \approx \chi \, u_{\rm iso}$ which corresponds to isoviscous flow with viscosity $\mu$.

 The volume rate (per unit length in the $z$ direction)
 \begin{equation}
 Q = 2 \left[\int_{a/2}^{\xi} u_-(y) \, dy + \int_{\xi}^a u_+(y) \, dy \right]
 \end{equation}
Substituting for the velocities and evaluating yields
\begin{equation}
\frac{Q \hfill}{Q_{\rm iso}} = \left(2 \, \chi - 1 \right)  - 2 \, (\chi-1) \, \frac{\xi^2}{a^2} \left(3 - 2 \, \frac{\xi}{a} \right) 
\end{equation}
where
\begin{equation}
Q_{\rm iso} = \frac{\nabla p \, a^3}{12 \eta}
\label{qplan}
\end{equation}
Checking the isoviscous limit $\eta = \mu$, $\chi = 1$, yields $Q = Q_{\rm iso}$.  Likewise, for $\tau_c = \tau(a)$, $\xi = a$, and  $Q = Q_{\rm iso}$.  Finally, in the large strain rate limit  $\tau_c \ll \tau(a)$, $\xi \rightarrow a/2^+$, and $Q \approx \chi \, Q_{\rm iso}$. 

\subsection{Cylindrical geometry}

Consider cylindrical conduit geometry with axial coordinate $z$ and no slip surface at radius $r = R$. Flow is driven by a uniform pressure gradient, $p(z) = - \nabla p \, z$ with $\nabla p > 0$.  In this case, the steady state velocity $\boldsymbol{u} = u(r) \, \hat{\boldsymbol{z}}$. 

The non-zero components of the stress tensor
\begin{equation}
\tau_{rz} = 2 \, \eta_{\rm eff} \, \dot{\varepsilon}_{rz} = 2 \, \eta_{\rm eff} \, \left(\frac{1}{2} \frac{d u}{d r}\right) = \eta_{\rm eff} \, \frac{d u}{d r}  
\end{equation}
The momentum equation reduces to
\begin{equation}
\frac{d p}{d z} = -\nabla p = \frac{1}{r} \frac{d \hfill}{d r} \left( r \tau_{rz} \right)
\end{equation}
Integrating and requiring that $\tau_{rz}$ be finite at $r=0$ yields
\begin{equation}
\tau_{rz} = - \frac{\nabla p}{2} \, r
\label{taurz}
\end{equation}
The stress invariant $\tau(r) = \nabla p/2 \, r$ is an increasing linear function of radial position with maximum at the conduit boundary $\tau(R) = \nabla p/2 \, R$.  For $\tau(R) < \tau_c$, there is a radial position $\xi < R$ where the stress invariant equals the rheological transition stress,
\begin{equation}
\xi = \frac{2 \, \tau_c}{\nabla p}
\label{xicyl}
\end{equation}

For $r < \xi$, the effective viscosity $\eta_{\rm eff} = \eta$.  As in the planar geometry case, denote the velocity in this region as $u_-$.  Then, substituting for $\tau_{rz}$ in Eq.\ (\ref{taurz}) gives 
\begin{equation}
\eta \, \frac{d u_-}{d r \hfill} = - \frac{\nabla p}{2} \, r
\end{equation} 
Integrating,
\begin{equation}
u_- = - \frac{\nabla p \, R^2}{4 \, \eta} \, \left( \frac{r}{R} \right)^2 + c_1
\end{equation}
where $c_1$ is a constant. 
For $r > \xi$, the effective viscosity is given by Eq.\ (\ref{eq:rheo}) with strain rate invariant $\frac{1}{2} \, |du_+/dr|$.  Letting $u_+$ be the velocity in the region $\xi \le r \le R$, Eq.\ (\ref{taurz}) becomes
\begin{equation}
\frac{\eta}{\chi} \left[ \frac{\tau_c}{\eta \left|\dfrac{d u_+}{d r \hfill} \right|} \left(\chi-1\right)+1\right] \dfrac{d u_+}{d r \hfill} = - \frac{\nabla p}{2} \, r
\end{equation} 
The velocity gradient is a monotonically decreasing function of $r$, i.e., $|du_+/dr| = - du_+/dr$.  Making this substitution, eliminating $\tau_c$ using Eq.\ (\ref{xicyl}), integrating, and applying the no slip boundary condition $u_+(R) = 0$ yields
\begin{equation}
u_+ = \frac{\nabla p\, R^2}{4 \, \eta} \, \left[\chi \left(1 - \frac{r^2}{R^2}\right) - 2 \, 
(\chi-1) \, \frac{\xi}{R} \, \left( 1 - \frac{r}{R} \right) \right]    
\end{equation}
Matching solutions $u_-(\xi) = u_+(\xi)$ gives the constant $c_1$, 
\begin{equation}
u_- = \frac{\nabla p\, R^2}{4 \, \eta} \, \left[ \chi + (\chi-1) \left( \frac{\xi^2}{R^2} - 2 \frac{\xi}{R} \right) - \frac{r^2}{R^2} \right] 
\end{equation}

In the isoviscous limit $\eta = \mu$ and $\chi =1$, $u_+=u_-=u_{\rm iso}$ where the isoviscous velocity
\begin{equation}
u_{\rm iso} = \frac{\nabla p \, R^2}{4 \, \eta} \left(1 - \frac{r^2}{R^2} \right) 
\end{equation}
When $\tau_c = \tau(R)$, $\xi=R$, the entire conduit is in the high viscosity regime, and $u_-=u_{\rm iso}$.  In the large strain rate limit, $\tau_c \ll \tau(R)$ and $\xi \ll R$.  Apart from a small strain rate region around $r=0$, the rest of the fluid is in the low viscosity regime.    In this case, $u_+ \approx \chi \, u_{\rm iso}$ which corresponds to isoviscous flow with viscosity $\mu$.

\begin{figure}[ht!]
\centering
\includegraphics{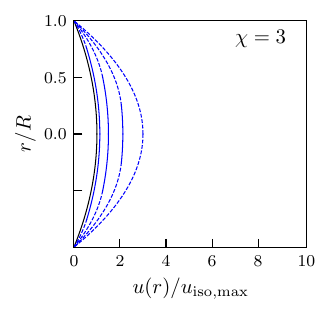}
\hspace{-0.25in}
\includegraphics{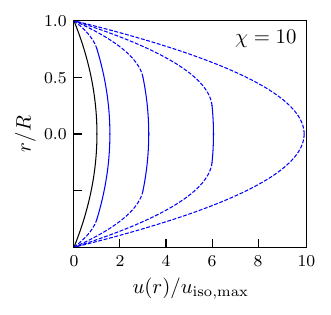}
\hspace{-0.25in}
\includegraphics{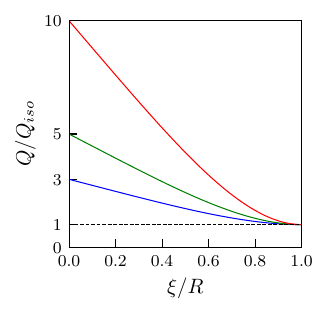}
\caption{Velocity profile dependence on viscosity ratio $\chi$ and rheological transition level $\xi$. The velocity is scaled by the maximum isoviscous velocity $u_{\rm iso,max} = (\nabla p \, R^2)/(4 \eta)$.  (left) The isoviscous case (solid black line) is compared to pseudoplastic solutions (blue lines) for $\chi=3$ and $\xi$ = (0.75, 0.50, 0.25, 0.00).  Layer regions where $\tau > \tau_c$ are denoted by dashed blue lines.  (center) Velocity profiles for $\chi$ = 10 and the same values of $\xi$.  (right) Volume rate as a function of rheological transition radius $\xi$. The isoviscous case (dashed black line) is compared to pseudoplastic solutions for $\chi$ = (3,5,10) shown in (blue, green, red) respectively.}
\label{fig3} 
\end{figure}

The volume rate 
\begin{equation}
Q = 2 \pi \left[ \int_0^{\xi} r \, u_-(r) \, dr + \int_{\xi}^R r \, u_+(r) \, dr \right] 
\end{equation}
Substituting for the velocities and integrating yields
\begin{equation}
\frac{Q \hfill}{Q_{\rm iso}} = \chi  - \frac{4}{3} \, (\chi-1) \, \frac{\xi}{R} \left(1 -  \frac{1}{4} \, \frac{\xi^3}{R^3} \right) 
\label{qcyl}
\end{equation}
where
\begin{equation}
Q_{\rm iso} = \frac{\pi}{8} \, \frac{\nabla p \, R^4}{\eta}
\end{equation}
Checking the isoviscous limit $\eta = \mu$, $\chi = 1$, yields $Q = Q_{\rm iso}$.  Likewise, for $\tau_c = \tau(R)$, $\xi = R$, and  $Q = Q_{\rm iso}$.  Finally, in the large strain rate limit  $\tau_c \ll \tau(R)$, $\xi \rightarrow 0$, and $Q \approx \chi \, Q_{\rm iso}$. 
\begin{figure}[ht!]
\centering
\includegraphics{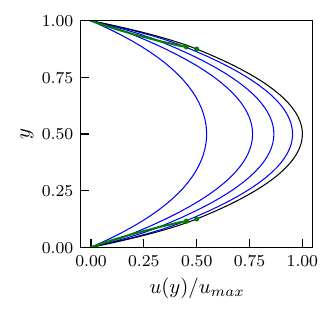}
\hspace{-0.25in}
\includegraphics{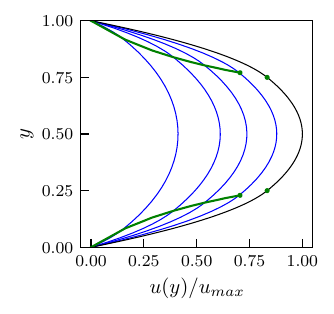}
\hspace{-0.25in}
\includegraphics{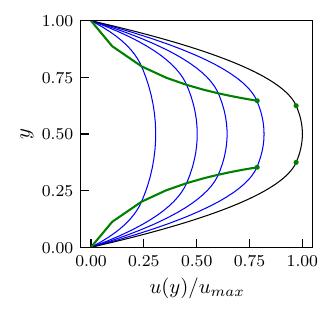}

\includegraphics{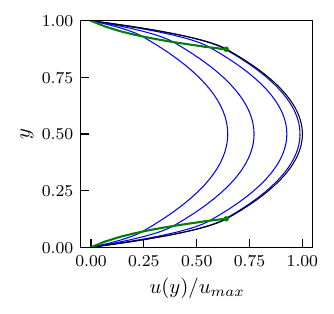}
\hspace{-0.25in}
\includegraphics{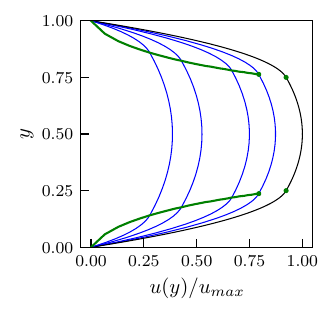}
\hspace{-0.25in}
\includegraphics{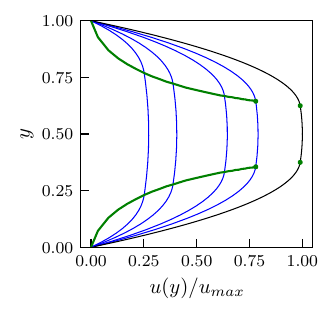}
\caption{Velocity evolution (blue) and steady state velocities (black) in planar geometry.  Velocities are scaled by the maximum steady state value $u_{\rm max } = 1/8 \, [1 + 4 (\chi-1)(1-\xi)^2]$.  The location of the rheological transition 
$\dot{\varepsilon} = \dot{\varepsilon}_c$ is shown in green (bold line/dots).  (top)  Velocity profiles for  $\chi=3$, steady state values $\xi$ = 0.825, 0.750, 0.625 (left, middle, right), and times $t$ = 0.1, 0.2, 0.3, 0.5.  (bottom)  As for the top with $\chi=10$ and times $t$ = 0.3, 0,5, 1.0, 1.5.}
\label{figu_plan} 3
\end{figure}

\section{Transient flow}

In this section, numerical simulations of flow development from a static initial state are considered.  The numerical 
method is benchmarked against isoviscous theory for planar and cylindrical geometries (see \S6). 
The approach to steady state is compared to theoretical results developed in \S3.  In particular, 
velocity distribution, extent of shear thinning, and volume rate are considered. 
\begin{figure}[ht!]
\centering
\includegraphics{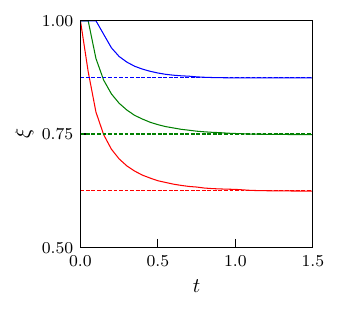}
\includegraphics{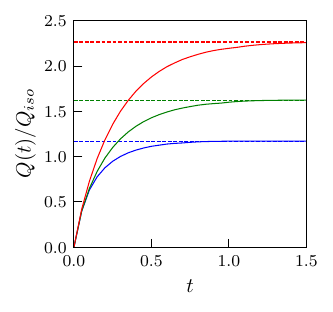}

\includegraphics{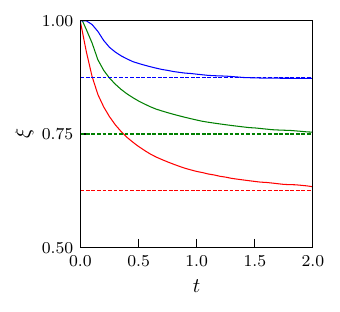}
\includegraphics{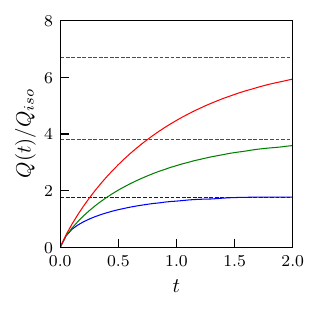}
\caption{(top) Evolution of the rheological transition location (left) and volume rate (right) for $\chi$ = 3.  The three cases correspond to steady state values $\xi$ = 0.825, 0.750, 0.625 (blue, green, red).  The volume rate is normalized by the isoviscous value based on $\eta$ (Eq.\ \ref{qplan}).  Steady values are shown with dashed lines.  (bottom) As for the top with $\chi$ = 10. }
\label{figxi_plan} 
\end{figure}     

\subsection{Planar geometry}

For the unidirectional flow considered here, the velocity gradient is perpendicular to the velocity resulting in a vanishing advective derivative $\boldsymbol{u} \cdot \grad \boldsymbol{u}$.  Substituing for $\tau_{xy}$ and pressure $p(x)$ in the momentum equation Eq.\ (\ref{cauchy}),
\begin{equation}
\frac{\partial u}{\partial t} = \frac{\partial}{\partial y} \left( \nu_{\rm eff}(\dot{\varepsilon}) \, \frac{\partial u}{\partial y} \right) + \frac{1}{\rho} \, \nabla p
\end{equation}
where $\nu_{\rm eff}(\dot{\varepsilon}) = \eta_{\rm eff}/\rho$ is the effective kinematic viscosity.  Rewriting $\eta_{\rm eff}$ in terms of the viscosity ratio $\chi$,
\begin{equation}
\nu_{\rm eff}(\dot{\varepsilon}) = \begin{cases}
\nu & \dot{\varepsilon}  \le   \dot{\varepsilon}_{\rm c} \\
\dfrac{\nu}{\chi} \left[ \dfrac{\tau_c}{2 \, \eta \, \dot{\varepsilon}} \left(\chi-1\right) +1 \right] & \dot{\varepsilon}  \ge   \dot{\varepsilon}_{\rm c} \\
\end{cases}
\label{nu}
\end{equation}  
where $\nu = \eta/\rho$.  
Non-dimensionalization with the following scales
\begin{equation}
[y] = a \qquad  [t] = \frac{a^2}{\nu} \qquad [u] = \frac{\nabla p \, a^2}{\rho \,  \nu} \qquad [\dot{\varepsilon}] = \frac{\nabla p \, a}{\rho \, \nu} \qquad [\tau] = \nabla p \, a
\end{equation}
yields the momentum equation
\begin{equation}
\frac{\partial u}{\partial t} = \frac{\partial}{\partial y} \left( \nu_{\rm eff}^{\prime}(\dot{\varepsilon}) \, \frac{\partial u}{\partial y} \right) + 1
\label{tran_plan}
\end{equation}
where
\begin{equation}
\nu_{\rm eff}^{\prime}(\dot{\varepsilon}) = \begin{cases}
1 & \dot{\varepsilon}  \le   \dot{\varepsilon}_{\rm c} \\
\dfrac{1}{\chi} \left[ \dfrac{\tau_c}{2 \, \dot{\varepsilon}} \left(\chi-1\right) +1 \right] & \dot{\varepsilon}  \ge   \dot{\varepsilon}_{\rm c} \\
\end{cases}
\label{nu_nondim}
\end{equation}  
The non-dimensional, transition stress invariant $\tau_c = \xi - 1/2$.  The numerical computation was initialized with a small
parabolic velocity profile $u(y,0) = \epsilon (y-y^2)$ with $\epsilon$ = 10$^{-4}$.  The spatial discretization interval $\delta y$ = 0.05.

Channel velocity evolution for $\chi$ = 3,10 and $\xi$ = 0.625, 0.750, 0.825 is shown in Fig.\ (\ref{figu_plan}).  
Depending on $\chi$ and $\xi$, steady state is approached after non-dimensional times $\sim$ 0.5-2.  
For large rheological transition stress, the fluid is essentially isoviscous with $\eta_{\rm eff} \sim \eta$ and 
steady state is reached after non-dimensional time $\sim$ 0.5. 
When the bulk of the fluid is in the low viscosity regime (small $\xi$) and $\chi$ is large, the 
flow adjustment time is longer.  
In the limit of fully developed shear thinning ($\xi \rightarrow$ 0), the momentum diffusion time is governed 
by  $\eta_{\rm eff} \sim \mu$ and time to reach steady state scales linearly with $\chi$.  
Pseudoplastic transition location and volume rate as a function of time are shown in Fig.\ (\ref{figxi_plan}).  
\begin{figure}[ht!]
\centering
\includegraphics{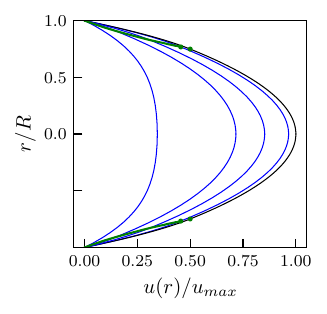}
\hspace{-0.25in}
\includegraphics{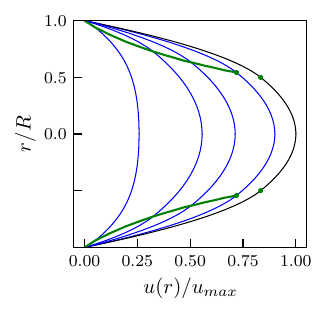}
\hspace{-0.25in}
\includegraphics{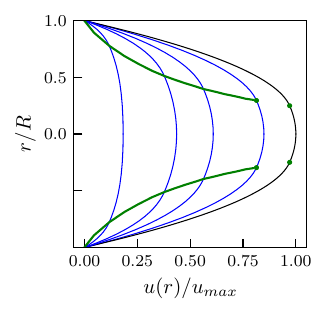}

\includegraphics{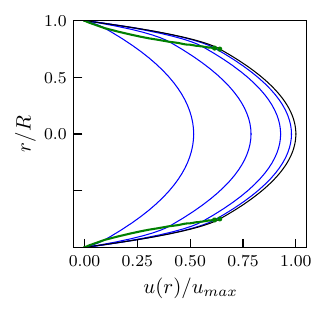}
\hspace{-0.25in}
\includegraphics{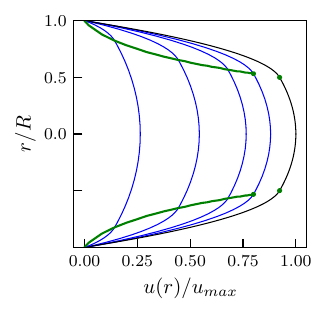}
\hspace{-0.25in}
\includegraphics{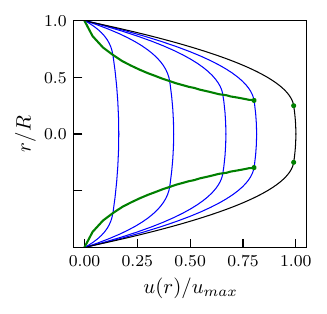}
\caption{Velocity evolution (blue) and steady state velocities (black).  Velocities are scaled by the maximum steady state value $u_{\rm max }=1/4\, [\chi - (\chi-1)\, \xi\, (2-\xi)]$.  The location of the rheological transition $\dot{\varepsilon} =\dot{\varepsilon}_c$ is shown in green (bold line/dots).  (top)  Velocity profiles for  $\chi=3$, steady state values $\xi$ = 0.75, 0.50, 0.25 (left, middle, right), and times $t$ = 0.1, 0.3, 0.5, 1.0.  (bottom)  As for the top with $\chi=10$ and times $t$ = 0.3, 1.0, 2.0, 3.0.}
\label{figu_cyl} 
\end{figure}

\subsection{Cylindrical geometry}
In cylindrical geometry, the transient momentum equation is
\begin{equation}
\frac{\partial u}{\partial t} = \frac{1}{r} \, \frac{\partial}{\partial r} \left(r \, \nu_{\rm eff}(\dot{\varepsilon}) \, \frac{\partial u}{\partial r} \right) + \frac{1}{\rho} \, \nabla p
\end{equation}
with $\nu_{\rm eff}$ given by Eq.\ (\ref{nu}).  
Non-dimensionalizing with the following scales
\begin{equation}
[r]=R \qquad [t]=\frac{R^2}{\nu} \qquad [u]=\frac{\nabla p \, R^2}{\rho \, \nu} \qquad [\dot{\varepsilon}] = \frac{\nabla p \, R}{\rho \, \nu} \qquad [\tau] = \nabla p \, R
\end{equation}
yields 
\begin{equation}
\frac{\partial u}{\partial t} = \frac{1}{r} \, \frac{\partial}{\partial r} \left(r \, \nu_{\rm eff}^{\prime}(\dot{\varepsilon}) \,  \frac{\partial u}{\partial r} \right) + 1
\label{tran_cyli}
\end{equation}
where $\nu_{\rm eff}^{\prime}$ is given by Eq.\ (\ref{nu_nondim}).  The non-dimensional, rheology transition stress invariant $\tau_c = \xi/2$.  
The computation was initialized with a small parabolic velocity profile satisfying the no slip boundary condition, $u(r,0) = \epsilon (1-r^2)$ with $\epsilon$ = 10$^{-4}$.  The spatial discretization interval $\delta r$ = 0.05.

\begin{figure}[ht!]
\centering
\includegraphics{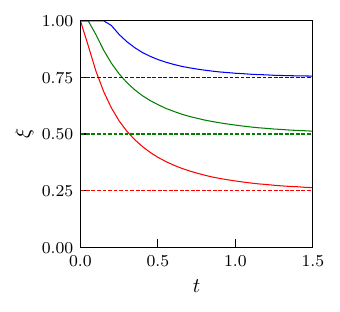}
\includegraphics{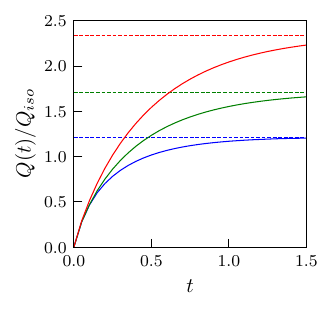}

\includegraphics{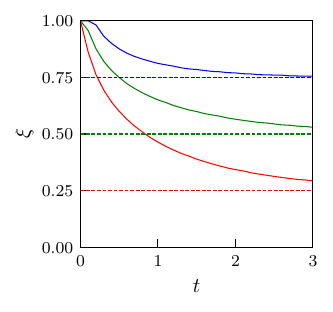}
\includegraphics{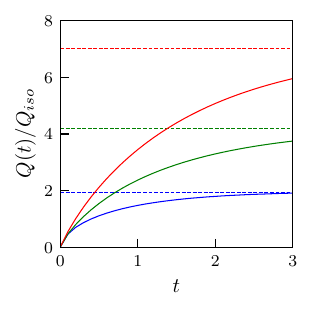}
\caption{(top) Evolution of the rheological transition location (left) and volume rate (right) for $\chi$ = 3.  The three cases correspond to steady state values $\xi$ = 0.75, 0.50, 0.25 (blue, green, red).  The volume rate is normalized by the isoviscous value based on $\eta$ (Eq.\ \ref{qcyl}).  Steady values are shown with dashed lines.  (bottom) As for the top with $\chi$ = 10. }
\label{figxi_cyl} 
\end{figure}

Velocity evolution for $\chi$ = 3,10 and $\xi$ = 0.75, 0.50, 0.25 is shown in Fig.\ (\ref{figu_cyl}) for pipe flow.  Velocities approach steady state after non-dimensional times $\sim$ 1-3 depending on spatial extent of shear thinning and viscosity contrast.  
For sufficiently large $\tau_c$, the fluid is essentially isoviscous with $\eta_{\rm eff} \sim \eta$ and the 
momentum diffusion time is $\sim$ 1.
When more of the fluid is in the low viscosity regime (small $\xi$) and the viscosity contrast $\chi$ is large, the flow adjustment time is longer.  
In the limit of fully developed shear thinning ($\xi \rightarrow$ 0), momentum diffusion is governed by  $\eta_{\rm eff} \sim \mu$ and the flow adjustment timescale  $\sim \chi$.  
Evolution of pseudoplastic transition location and volume rate are shown in Fig.\ (\ref{figxi_cyl}).  

\section{Discussion}

In this paper, pseudoplastic fluid flow in flat, channel and pipe geometries is considered.  
A simplified model which captures fundamental characteristics of more complicated 
shear thinning fluid rheologies is adopted.  The model admits closed form solutions for
steady state velocity distributions expressed as piecewise polynomial functions in spatial 
coordinates.  Likewise, volume rates are simple analytical functions of rheological parameters and
driving pressure gradient.
Analytical simplicity of the results may prove useful for parameter space exploration prior to 
adoption of more sophisticated rheologies and/or appeal to finite element simulations.

Transient flow evolution from a static state is investigated via numerical computation.  
It is found that model fluid flow approaches theoretical steady state on timescales 
governed by the spatially dominant effective kinematic viscosity.  That is to say, the 
reduced effective viscosity governs stationary flow development
timescale when shear thinning occurs throughout the majority of the fluid.  
In the steady state limit, numerical results are in good agreement with theory.  

\begin{figure}[ht!]
\centering
\includegraphics{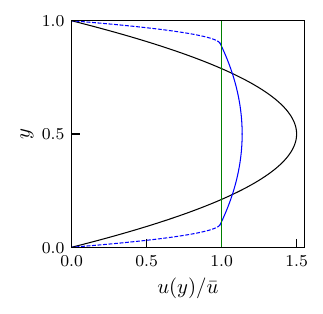}
\caption{Planar, non-dimensional velocity profiles for isoviscous (black) and pseduoplastic (blue) 
cases with equal average (plug) velocity $\bar{u}$.  The pseudoplastic 
case has $\chi$ = 100 and $\xi$ = 0.9 and  shear thinning regions are denoted by dashed lines.}
\label{fig_plug} 
\end{figure}

For large viscosity ratio $\chi$ and shear thinning regions confined near no-slip boundaries
the flow can exhibit plug-like characteristics.  
This is illustrated in Fig.\ \ref{fig_plug} for a plane layer with shear thinning case 
$\chi$ = 100 and $\xi$ = 0.9 such that 
narrow, large strain-rate layers exist adjacent to domain boundaries.
The velocity profile is compared to the isoviscous case for equal average velocities.   

The simplified rheology and results considered in this work  may find application in 
models of dike emplacement and/or magma 
flow in volcanic conduits \citep{gonnermann07} as silicic magmas 
exhibit shear thinning behavior \citep{lavallee07,jones20,vasseur23}.  
The model  may also prove useful in industrial applications 
when volume rate control of generalized Newtonian fluids is of
interest (e.g., in the case of polymeric liquids \citep{bird87}).     
       
\clearpage       
\section{Appendix: Numerical methods}

In this appendix, the numerical method is benchmarked against isoviscous analytical solutions for transient development of Poiseuille flow in planar and cylindrical geometries from zero velocity to 
steady state.  The transient momentum equations (Eqs.\ \ref{tran_plan}, \ref{tran_cyli}) are of the form
\begin{equation}
\frac{\partial u(r,t)}{\partial t} = \mathcal{D}\left[u(r,t)\right]
\end{equation}
where $\mathcal{D}$ is a non-linear differential operator.  The Python package py-pde \citep{zwicker20} provides methods for solving partial differential equations of this form.  The py-pde, method-of-lines scheme utilizes implicit, Adams backward differentiation \citep{hindmarsh83,petzold83}. 

\subsection{Planar geometry}
Consider flow start-up from an initial zero velocity state $u(y,0) = 0$ subject to no slip conditions at the channel boundaries, $u(0,t) = u(1,t) = 0$.
The non-dimensional, analytical solution of the inhomogeneous diffusion equation (Eq.\ \ref{tran_plan}) 
with $\nu^{\prime}_{\rm eff} = 1$ subject to these initial and boundary conditions is
\begin{equation}
u(y,t) = 4 \, \sum_{n \; {\rm odd}} \frac{\sin n \pi y}{(n \pi)^3} \left(1 -  e^{-n^2 \pi^2 t} \right) \; .
\end{equation}
Integrating the velocity profile over the layer thickness gives the volume rate
\begin{equation}
Q(t) = 8 \, \sum_{n \; {\rm odd}} \frac{1 - e^{-n^2 \pi^2 t}}{(n \pi)^4} \; .
\end{equation}
For comparison to computational results, sums were truncated at 11 non-zero terms. For non-dimensional time $t$ = 1, the next series term for the volume rate is $\sim$ 10$^{-5}$ times 
the partial sum.  

\begin{figure}[ht!]
\centering
\includegraphics{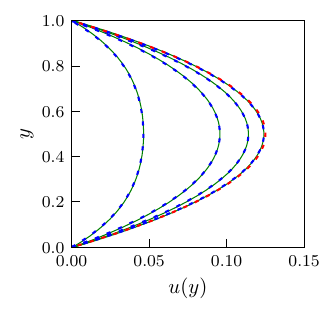}
\hspace{0.25in}
\includegraphics{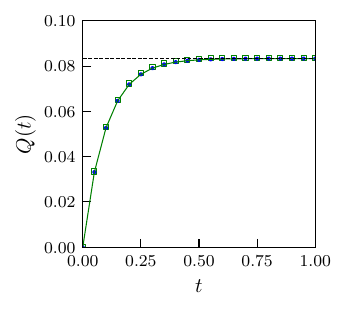}
\caption{Benchmark for startup of isoviscous planar Poiseuille flow. (left) Numerical (green solid line) and analytical (blue dashed line) velocity profiles at non-dimensional times 0.05, 0.15, 0.25, and 0.50.  Solutions approach the steady state analytical profile (red dashed line) at time $t \sim$ 0.5.  (right)  Numerical (green squares/line) and analytical (blue circles) transient volume rate.  The non-dimensional steady state value $Q =$ 1/12 is shown with a dashed black horizontal line.}
\label{plan_bench} 
\end{figure}

Discretization and initial condition are the same as for the pseudoplastic computations (\S4.1). Numerical and analytical results are in good agreement during development of steady state flow (Fig.\ \ref{plan_bench}).  Velocity distribution and volume flux solutions approach steady state solutions at time $t \sim$ 0.5.  For time $t$ = 1.0 the absolute error between analytical and numerical velocity profiles integrated over the layer is $\sim$ 10$^{-5}$.

\subsection{Cylindrical geometry}
For initial condition  $u(r,0)=0$ and no slip boundary condition $u(1,t)=0$,  
the non-dimensional, analytical velocity profile 
\begin{equation}
u(r,t) = \frac{1}{4} \left(1 - r^2\right) - 2 \, \sum_n \frac{J_0(\lambda_n r)}{\lambda_n^3 \; J_1(\lambda_n)} e^{-\lambda_n^2 t} \; ,
\end{equation}
where $J_0$ and $J_1$ are Bessel functions of the first kind of order zero and one respectively and 
$\lambda_n$ are the positive roots of $J_0$. 
The associated volume rate
\begin{equation}
Q(t) = \frac{\pi}{8} - 4 \pi \sum_n \frac{e^{- \lambda_n^2 t}}{\lambda_n^2}
\end{equation}
For comparison to computational results, sums were truncated at 10 terms. For non-dimensional time $t$ = 1, the next series term for the volume rate is vanishingly small.  

\begin{figure}[hb!]
\centering
\includegraphics{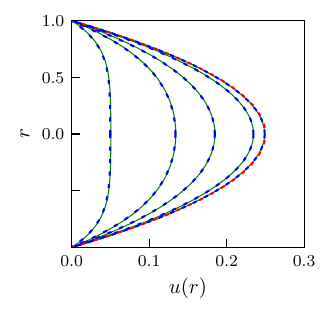}
\hspace{0.25in}
\includegraphics{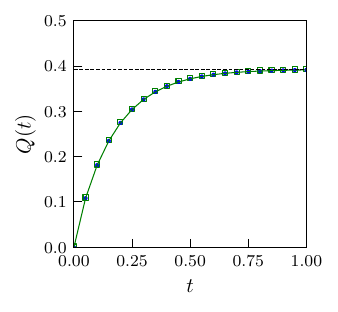}
\caption{Benchmark for startup of isoviscous cylindrical Poiseuille flow. (left) Numerical (green solid line) and analytical (blue dashed line) velocity profiles at non-dimensional times 0.05, 0.15, 0.25, 0.50, and 1.0.  Solutions approach the steady state analytical profile (red dashed line) at time $t \sim$ 1.0.  (right)  Numerical (green squares/line) and analytical (blue circles) transient volume rate.  The non-dimensional steady state value $Q = \pi/8$ is shown with a dashed black horizontal line.}
\label{cyli_bench} 
\end{figure}

Discretization and initial condition are the same as for the pseudoplastic computations (\S4.2).  Numerical and analytical results are in good agreement during development of steady state flow (Fig.\ \ref{cyli_bench}).  Velocity distribution and volume flux solutions approach steady state solutions at time $t \sim$ 1.0.  For time $t$ = 1.0 the absolute error between analytical and numerical velocity profiles integrated over the layer is $\sim$ 3 $\times$ 10$^{-4}$.

\end{document}